# Moderator Factors of Software Security and Performance Verification

**Victor Vidigal Ribeiro**
Federal University of Rio de Janeiro
P.O. Box 68.511
Rio de Janeiro, Brazil
vidigal@cos.ufrj.br

**Daniela Soares Cruzes**
SINTEF DIGITAL
P.O. NO-7465
Trondheim, Norway
danielac@sintef.no

**Guilherme Horta Travassos**
Federal University of Rio de Janeiro
P.O. Box 68.511
Rio de Janeiro, Brazil
ght@cos.ufrj.br

## ABSTRACT

**Context:** Security and performance are critical software non-functional requirements. Therefore, verification activities should be included in the development process to identify related defects, avoiding failures after deployment. However, there is a lack of understanding on factors moderating the security and performance verification, which jeopardizes organizations to improve their verification activities to assure the releasing of software fulfilling these requirements.

**Objective:** To identify moderator factors influencing security and performance verification and actions to promote them.

**Method:** Case study to identify security and performance moderators factors. Rapid Literature Reviews with Snowballing to strengthen moderator factors confidence. Practitioners Survey to classify the moderator factors relevance.

**Results:** Identification of eight security and performance moderator factors regarding organizational awareness, cross-functional team, suitable requirements, support tools, verification environment, verification methodology, verification planning, and reuse practices. Rapid Reviews confirmed the moderator factors and revealed actions to promote each. A survey with 37 practitioners allowed us to classify the moderator factors and their actions regarding their relevancy.

**Conclusions:** The moderator factors can be considered key points to software development organizations implement/improve security and performance verification activities in regular software systems. Further investigation is necessary to support the understanding of these moderator factors when building modern software systems.

**Keywords:** security, performance, software verification, software testing, evidence-based software engineering.

## 1 INTRODUCTION

The popularization and extensive use of software systems bring benefits to modern life. However, their full availability and connectivity increase specific concerns regarding critical software quality dimensions, including security (Ameller et al., 2016) and performance (Caracciolo et al., 2014). Security is relevant owing to vital and sensitive information manipulated and stored by the software systems while performing their tasks. For instance, software systems are responsible for managing personal data, organizations' strategic communication, and financial transactions control. This information usually requires high confidence and different classification levels, resulting in a growing interest in obtaining improper benefits (Labs 2016) (Threat and Index 2017). Performance is relevant owing to the limitations of computational resources (Zhu et al., 2015). Long response time can make users migrate to rival software systems. A delayed financial transaction can result in financial losses. Excessive power consumption can make the use of systems unfeasible (if hosted on battery-based devices) or increase the energy costs of systems running in large data centers.

Software development organizations usually include quality assurance activities throughout the software life cycle to evaluate software quality, preventing software release failures. Software verification (IEEE-1012 2017), including testing and reviewing, encompasses various activities to analyze the software searching for defects. Security and Performance (S&P) verification are activities that search for defects regarding these specific quality perspectives. Different verification practices and techniques can be used individually or combined, promoting clear benefits but also posing various challenges to the verification of S&P (Atifi et al. 2017) (Felderer et al. 2016) (Meira et al. 2016).

Despite some S&P verification techniques, software systems still present several defects related to these quality properties. For example, performance issues account for a significant fault category in specific domains (*e.g.*, telecommunications) (Bertolino 2007), and news reporting systems attacks are increasingly frequent (Symantec 2017). It may be attributed to (1) the inefficiency of S&P verification practices, (2) the fact that software organizations do not adopt reasonable verification practices, or (3) the lack of evidence-based verification practices owing to the apparent disconnection between academy and industry in this context (Garousi and Felderer 2017). Furthermore, automated attack scripts, the abundance of attack information, and global interconnection make it easier to attack systems than previously (Vaughn et al., 2002).

Thus, in the current scenario, identifying the factors that influence S&P verification is essential. Furthermore, the factors work as a guide for software development organizations to know where to improve S&P verification activities and produce more security and better performance.

This paper presents eight moderator factors (MF) influencing S&P verification activities that contribute to the S&P verification topic. Initially, the MFs emerged from a case study performed on three organizations as conjectures[1] and reported in Ribeiro *et al.* (2018):

- Security and performance verification requires a suitable environment;
- Security and performance verification requires appropriate techniques;
- The lack of security and performance requirements prevents the guarantee from fulfilling its original purpose;
- Training in security and performance verification improves verification activities;
- Security and performance verification requires cross-functional teams;
- Environment configuration and planning of security and performance verification Demands extra effort that is not always accounted for;
- Organizational support influences security and performance activity positively;
- Security and performance verification activities are ignored when there are limited resources of time and budget;
- Verification activities alone do not guarantee software security and performance effectiveness.

Realizing the importance of the conjectures, we started a more in-depth investigation to understand whether such conjectures would also apply to a broader context. Thus, we performed the case study in a new organization and rearranged the conjectures to reflect the new findings. In sequence, we performed rapid reviews with snowballing, aiming to improve their confidence. After rearranging and confirming the conjectures through the technical literature, we reclassified them into eight moderator factors presented in this work. Additionally, we performed a survey aiming to identify the moderator factors' relevance according to practitioners.

In the remainder of the paper, we first introduce the main related concepts, aiming to bring the readers into our perspective and improve understanding. Then, section 3 presents the research methodology, providing important context information. Next, section 4 presents the identified S&P moderator factors and the actions used to promote them. Section 5 sets the work concerning existing literature, presenting the related works. Section 6 presents the threats to validity. Section 7 discusses the results and exposes questions that remain open. Finally, Section 8 concludes the paper.

---

[1] Inference formed statements without proof or sufficient evidence ("Conjecture" 2011)

## 2  BACKGROUND

First, it is vital to justify the use of two non-functional requirements (NFRs): security and performance. The research starting point was originally to aid software developers in the verification of any non-functional requirements. Thus, the first research activity was to identify the relevant NFRs, resulting in more than two hundred different NFRs (Ribeiro and Travassos 2016); however, it was unfeasible to handle all these NFRs a single study. Accordingly, we chose security and performance, as they are cited as essential NFRs in technical literature and various software organizations. Furthermore, even though a single NFR could have been chosen, the software has multiple quality dimensions, and it is crucial to understand their inherent trade-offs. Therefore, these two most essential NFRs were selected.

Overall, this study follows the concepts defined by ISO-29119 (2013). Verification encompasses activities for evaluating a software system (or a part of it) to determine whether the developed product satisfies the requirements. The verification activities can be classified into dynamic (the software must be running, i.e., software testing) and static (the software is not running, i.e., software reviews).

Subsequently, the most important concepts discussed in this study are briefly highlighted, with no intention of re-conceptualizing all software verification areas (Table 1).

**Table 1** Related concepts

| Concept | Interpretation |
|---|---|
| **Asset** | The system covered by the verification practice, e.g., the source code, is an asset regarding static code analysis. However, it is not defined as an artifact because the verification may target the running system, which is not an artifact. |
| **Automation level** | When looking at the automation level, we need to know if a practice is performed manually or automated. |
| **Defect** | It is a general concept for representing a Failure or Fault. The term defect is used in this work since it is more generic and can also represent security and performance issues. |
| **Definition of done, acceptance criteria, or stop criteria** | Overlapping concepts. The definition of done is used as a criterion to conclude a verification activity. |
| **Failure** | The inability of a system or component to perform its required functions. The manifestation of a Fault. |
| **Fault** | It represents an incorrect step, process, or data definition in a computer program. |
| **Performance verification** | Verification activities aim to identify defects affecting software performance. |
| **Security verification** | Verification activities aim to identify defects affecting software security. |
| **Security and performance (S&P) moderator factors (MFs)** | Important dimensions that should be observed when handling security and performance verification. |
| **Verification practice** | <u>What</u> is performed for supporting verification, e.g., unit testing. |
| **Verification technique** | <u>How</u> the verification is performed (e.g., boundary value to generate test cases). |
| **Vulnerability** | It represents a security fault (or defect). |

## 3  OVERALL STUDY METHODOLOGY

This study is a follow-up from a case study performed on four different organizations. It was conducted to identify how S&P verification is performed and how decision-making regarding S&P verification is made in software development organizations. This study's findings were organized as a set of conjectures about S&P verification and presented in Ribeiro *et al.* (2018). Given the importance of these conjectures, a group of studies were conducted in the form of rapid reviews (RRs) (Tricco et al. 2015) and snowballing (Wohlin 2014a) to search for support literature to confirm the findings. After analyzing the extraction forms of the RRs through a coding process and thematic analysis, it was possible to consider the conjectures as moderator factors influencing the verification of S&P. It was also possible to identify actions to promote such S&P moderator factors. Finally, a survey was performed to identify the relevance of S&P moderator factors (MFs) and their actions, evidencing they are relevant according to practitioners' opinions. Figure 1 shows an overview of the methodology and the replication package is published in Ribeiro *et al.* (2021).

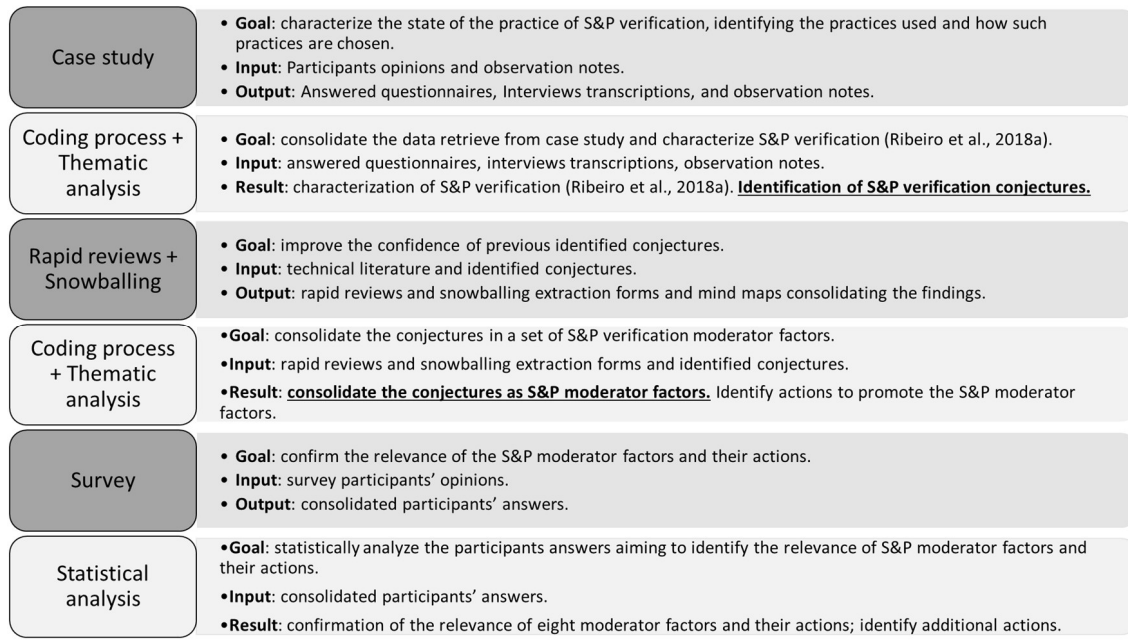

**Figure 1** General research methodology

## 3.1 CASE STUDY METHODOLOGY

The case study follows the recommendations presented in the guidelines proposed by Runeson and Höst (2009). Table 2 shows the mains sections of its protocol (Ribeiro et al. 2021).

**Table 2** Case study research protocol

| **Objectives** |
|---|
| Understand how security and performance verification has been performed in the software development industry. |

| **Scope** |
|---|
| Software development organizations performing Security OR Performance Verification activities. |

| **Research method** |
|---|
| <ul><li>Multiple case studies:<ul><li>One organization with one project as the primary case: including observational, semi-structured, and questionnaires data collection method;</li><li>Three organizations with one project each as complementary cases: including semi-structured and questionnaires data collection method;</li></ul></li><li>Flexible design<ul><li>Protocol improvements during the study execution.</li></ul></li><li>Predominantly qualitative</li><li>Criteria for case selection<ul><li>Projects in progress for at least two months</li></ul></li></ul> |

| **Data sources** |
|---|
| Organization's employers, researcher observations, institutional websites |

| **Unit of analysis** |
|---|
| Software development projects, including S&P verification activities. |

### 3.1.1 Describing Organizations

Conceivably, the diversity of the organizations' profiles will facilitate the generalization of the results (Table 3).

**Table 3** Organizations' description

| Id | Nature | #Employees (#Developers) | #Subjects | Data collection method |
|------|------------------------------|---------------------|-----------|-----------------------------------------|
| **Org1** | Governmental | ~10599 (Unknown) | 5 | Observation; Interview; Questionnaire |
| **Org2** | University tech transfer laboratory | ~154 (~132) | 2 | Interview; Questionnaire |
| **Org3** | Private | ~250 (150) | 2 | Interview; Questionnaire |
| **Org4** | Military | ~80507 (Unknown) | 3 | Interview; Questionnaire |

Org1 is a substantial governmental organization with ~10599 employees that provides information technology services to the Brazilian government. Most of its employees are software developers. A university tech transfer lab represents Org2. It has approximately 154 employees (132 are software developers) and develops technical solutions for the Brazilian government, including software development services. Org3 is a private company that generates credit card payment software systems, with approximately 250 employees and 150 software developers. Finally, org4 is a military organization that made it possible to investigate a large development department dealing with safety-critical software.

The organizations were also characterized by their agility level (Table 4). Although not exhaustive, this characterization provides a perception of the practices adopted by organizations. Thus, the participants were asked about the used agile practices. Some participants reported that they failed to implement agile practices. For example, a participant from Org3 said it was challenging to keep frequent contact with the clients because of their availability. The difficulty in implementing an agile methodology in Org 4, a military organization, is related to its rigid employee hierarchy.

It is noted that automated testing and Scrum are adopted practices in all organizations. Continuous integration and Kanban are the next most used practices. It is advised that this classification does not confirm that one organization is more agile than another. The number of observations points could influence the number of agile practices implemented by an organization. Thus, such information aims to enrich the organization characterization, thus facilitating the interpretation of the study results in similar contexts.

**Table 4** Organizations' agility level

| Agile Practice | Org 1 | Org 2 | Org 3 | Org 4 |
|-------------------------------|-------|-------|-------|-------|
| **Automated testing** | X | X | X | X |
| **Continuous integration** | X | | X | X |
| **Frequent meets with the client** | | X | | |
| **Internal daily meetings** | X | | | |
| **Kanban** | X | | X | X |
| **Scrum-based** | X | X | X | X |
| **Self-allocation of tasks** | X | | | |
| **Squad-based** | | | X | |
| **Test-driven development** | | X | | |

### 3.1.2 Describing Participants

Data were collected from twelve participants (Table 5). It is important to note that the participants have a technical profile without executive power, and their experience is not accurate because it is self-reported. Furthermore, the symbol × implies that the participant did not answer the question.

Regardless of the participants' leading role (for instance, P1.4 is a software architect), all participants from Orgs 1, 2, 3, and P4.1 directly perform some activity related to security and/or performance verification. However, participants P4.2 and P4.3 do not directly perform verification activities, but they are involved indirectly by requesting and evaluating the output of such activities.

Participant P4.2 does not have experience in testing and NFR. However, s/he is part of a project that contemplates security and performance verification. Thus, this report is considered essential.

**Table 5** Participants' characterization

| Question | | Main role | Education level | Software development experience (months) | Software testing experience (months) | NFR testing experience (months) | # of software development projects |
|---|---|---|---|---|---|---|---|
| **Org 1** | P1.1 | Test analyst | Master D. | 144 | 96 | 12 | 20 |
| | P1.2 | Develop. analyst | Undergrad. | 168 | 96 | 12 | 12 |
| | P1.3 | Develop. analyst | Undergrad. | 144 | 6 | 6 | 20 |
| | P1.4 | Software architect | Lato sensu specializ. | 132 | 24 | 12 | 20 |
| | P1.5 | Test analyst | Lato sensu specializ. | 276 | × | × | 20 |
| **Org 2** | P2.1 | Security test analyst | Undergrad. | × | 84 | 48 | 70 |
| | P2.2 | Security test analyst | Undergrad. | × | 60 | 60 | × |
| **Org 3** | P3.1 | Programmer | Undergrad. | 120 | 48 | × | 20 |
| | P3.2 | Security test analyst | Lato sensu specializ. | × | × | 120 | × |
| **Org 4** | P4.1 | Security test analyst | Master D. | 360 | 12 | 12 | 15 |
| | P4.2 | Full-stack developer | Lato sensu specializ. | 120 | 0 | 0 | 20 |
| | P4.3 | Software architect | Lato sensu specializ. | 180 | 72 | 72 | 10 |

## 3.2 RAPID REVIEWS AND SNOWBALLING METHODOLOGY

Because the conjectures arose from practice and the participants' opinions, a set of rapid reviews (RR) were performed, consulting the technical literature, and searching for confirmations for these conjectures.

An RR is a secondary study aiming to promptly deliver evidence to practice with lower effort than a traditional systematic review. To be faster, RRs simplify some steps of systematic reviews. For instance, the database search is limited, the quality appraisal is eliminated, or only one researcher is used to analyze the collected data (Tricco et al., 2015).

Eight RRs were conducted following protocols available on Ribeiro *et al.* (2021). As the protocols followed the same template, here we present an example of the RR undertaken to confirm the conjecture "Security and performance verification requires a **suitable environment**." Table 6 shows the search string.

**Table 6** Rapid reviews search string example

| Template | ( "security verification" OR "performance verification" OR "security testing" OR "performance testing") AND (**environment**\*) AND ( "software" ) AND ( "benefit\*" OR "problem\*" OR "challenge\*" OR "strateg\*" OR "empirical stud\*" OR "experimental stud\*" OR "experiment\*" OR "case stud\*" OR "survey\*" ) |
|---|---|

The research questions followed the structure presented in Table 7.

**Table 7** Rapid reviews research questions example

| RQ 1 | What are the **benefits** of a **<u>suitable environment</u>** for verification of security and performance? |
|---|---|
| RQ 2 | Which **problems** are caused by an **<u>unsuitable environment</u>** for verification of security and performance? |
| RQ 3 | What are the **challenges** of creating a **<u>suitable environment</u>** for the verification of security and performance? |
| RQ 4 | What **strategies** support the precise definition of a **<u>suitable environment</u>** for the verification of security and performance? |

After searching the Scopus search engine, the first author read the title and abstract of the returned papers and classified them as included or excluded according to the criteria presented in Table 8. Then, aiming to avoid bias, the second and third authors reviewed the excluded papers, eventually suggesting inclusions.

**Table 8** Rapid reviews inclusion criteria example

| |
|---|
| **1.** The paper must be in the context of software engineering; and |
| **2.** The paper must be in the context of performance and/or security verification; and |
| **3.** The paper must report a study related to a <u>suitable environment</u> for security or performance verification activities; and |
| **4.** The paper must report an evidence-based study grounded in empirical methods (e.g., interviews, surveys, case studies, formal experiment, among others) or a proof of concept; and |
| **5.** The paper must provide data to answer at least one of the RR research questions. |
| **6.** The paper must be written in the English language. |

Subsequently, the first author extracted the data from the included papers using the form shown in Table 9.

**Table 9** Rapid review extraction form

| **\<paper_id\>** <br> \<paper_reference\> | |
|---|---|
| Description | A brief description of the study objectives |
| Study type | Type of experimental (or not) study performed |
| Benefits | Set of benefits of implementing the conjecture |
| Problems | Set of problems caused by the conjecture |
| Challenges | The challenges faced on contemplating the conjecture |
| Strategies | The strategies used to implement the conjecture |

After data extraction, the first author performed snowballing (Wohlin 2014b) to increase the literature coverage. The snowballing was backward with only one iteration, and the starter set was the included papers of the RRs. Table 10 presents the numbers of articles found, included through RRs, included through Snowballing and total included.

**Table 10** Number of papers retrieved from technical literature.

| Conjecture | # Found | # RR | # Snowballing | # Total |
|---|---|---|---|---|
| C01 | 63 | 2 | 0 | 2 |
| C02 | 42 | 3 | 0 | 3 |
| C03 | 129 | 6 | 8 | 14 |
| C04 | 185 | 12 | 5 | 17 |
| C05 | 117 | 3 | 1 | 4 |
| C06 | 77 | 4 | 9 | 13 |
| C07 | 41 | 3 | 0 | 3 |
| C08 | 11 | 2 | 0 | 2 |

Next, the first author imported the extraction forms in the MAXQDA tool and performed a coding and a thematic analysis process. Additionally, the second and third authors iteratively revised the generated codes in several meetings throughout the process. The RRs output is a set of mind maps with high-level themes representing the technical literature's findings.

Finally, the RRs' themes were matched to the case study's themes (previously created). Thereby, the findings that emerged from the practice state are supported by the findings extracted from the state of the art. Accordingly, credence was lent to the conjectures, turning them into moderator factors of S&P verification.

## 3.3 SURVEY METHODOLOGY

The objectives of the survey are twofold but using the same questions:

**Confirmation (control group)**: the first one aims to validate our understanding regarding the information supporting the moderator factors and the actions used to promote them. As this information was collected through the case study, people who participated in the case study were used as the subjects (control group). In this validation phase, it was possible to get answers from four out of 12 participants.

**Assess the relevance (practitioners' opinion)**: the second objective of the survey is to assess the moderator factor's relevance and the actions used to promote them. Thus, we presented the study to the software development community using blog posts, social networks, software testing communities, questions-answers services, e-mail groups, and direct private messages. Thus, the survey was available for one month (July 2019), including 37 valid answers from different participants.

The survey used the VAS scale (Wewers and Lowe 1990) to capture participants' opinions regarding the relevance of moderation factors and yes/no questions to capture participants' views regarding the relevance of actions. It is essential to mention that the VAS is used to capture participants' perceptions of a particular event. The VAS consists of a horizontal line with two anchor points – the left point indicates the observed event's absence and the right point of the observed event's total agreement. The results obtained using this type of scale are visual and should be reported using the same scale. **Figure 2** shows an example of a question aiming to identify the respondents' opinion about the relevance of a specific moderator factor (organizational awareness of the importance of security and performance) and the actions used to promote it. The characterization of the surveyed population and the statistical analysis of the survey data is available on Ribeiro *et al.* (2021).

**Figure 2** Example of a question of the survey

### 3.3.1 Survey participants characterization

The survey included a characterization section to get information about the participants and organizations they work. Such characterization is vital to understand the context the moderator factors can be applied.

For example, it was possible to obtain responses from 11 countries, but most (18) came from Brazil, and seven participants did not answer about the country they live in (Figure 3).

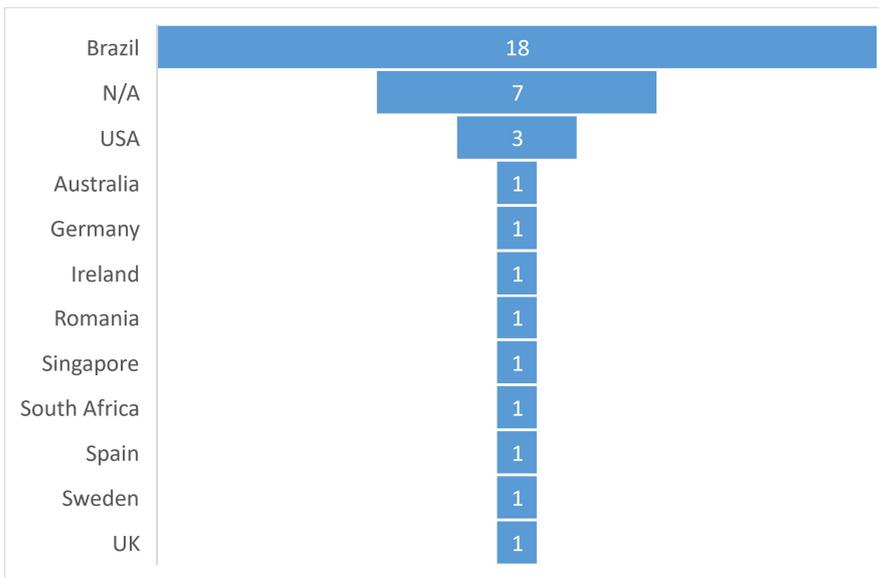

**Figure 3** Survey participants country

We were looking for practitioners involved in S&P verification activities, but they could play another role in the organization in which they work. Then we gather information about the primary role played by the participants. As shown in Figure 4, most of the participants are programmers, testers, or quality analysts.

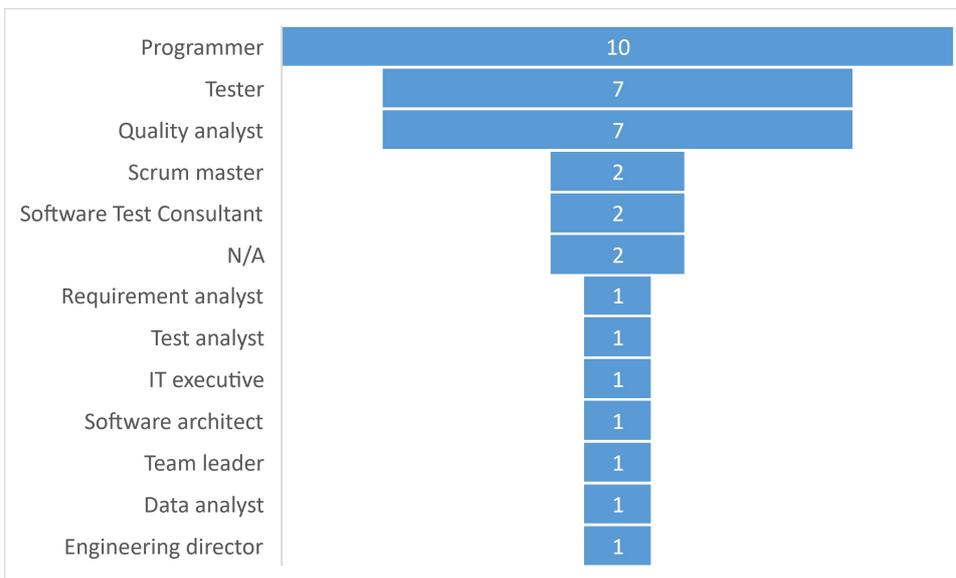

**Figure 4** Survey participants primary role

We also asked the participants for their experience in software development (Figure 5). Again, the results are divided according to the quartiles. Besides, five participants did not answer this question.

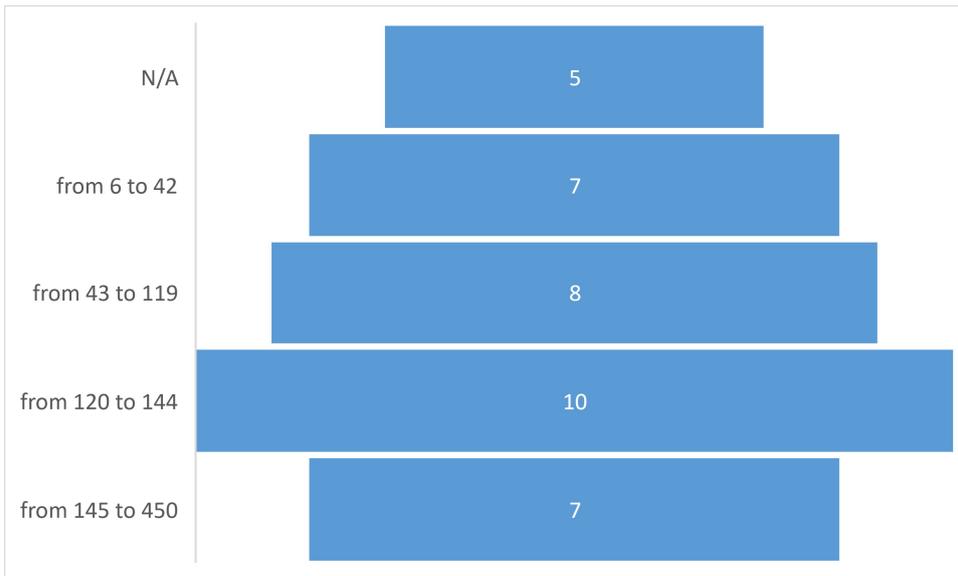

**Figure 5** Survey participants experience (months)

The size of the organizations (the number of employees) is also divided according to the quartiles. However, five participants did not answer the organization size (Figure 6).

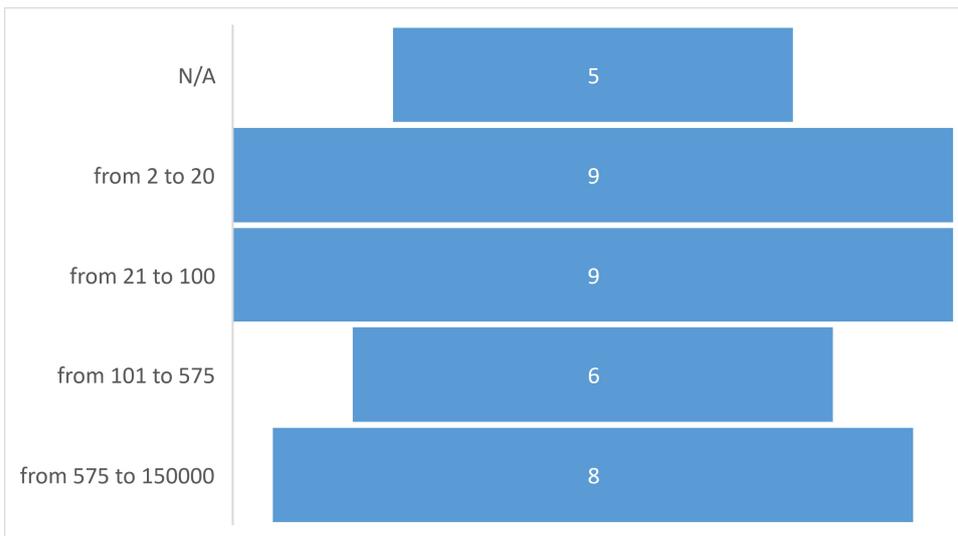

**Figure 6** Organizations' size (number of employees)

As shown in Figure 7, it was possible to collect participants' opinions working in organizations that develop software to distinct domains. The banking domain is the most frequent. It can indicate that the people working in this kind of organization care more about the S&P requirements. As some organizations develop software for different domains, the total is different from 37.

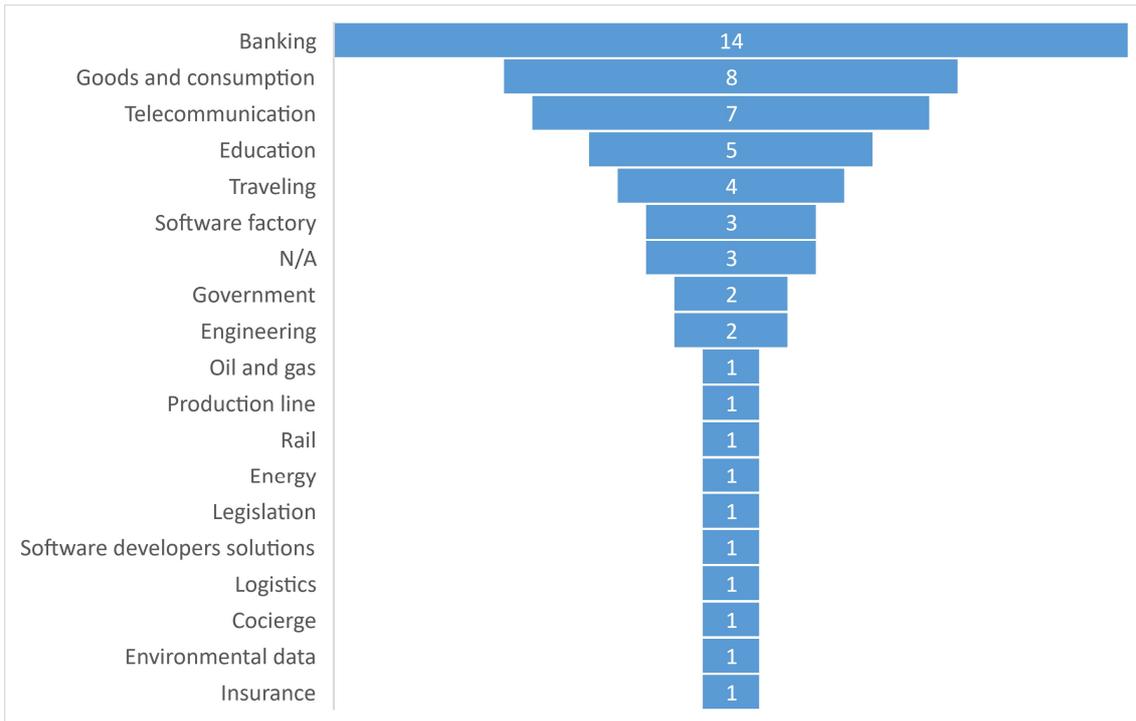

**Figure 7** Organizations' domain

Finally, we asked the participants regarding the agile practices used in their organizations. Such information can be used to understand better the context where the participant works (Figure 8).

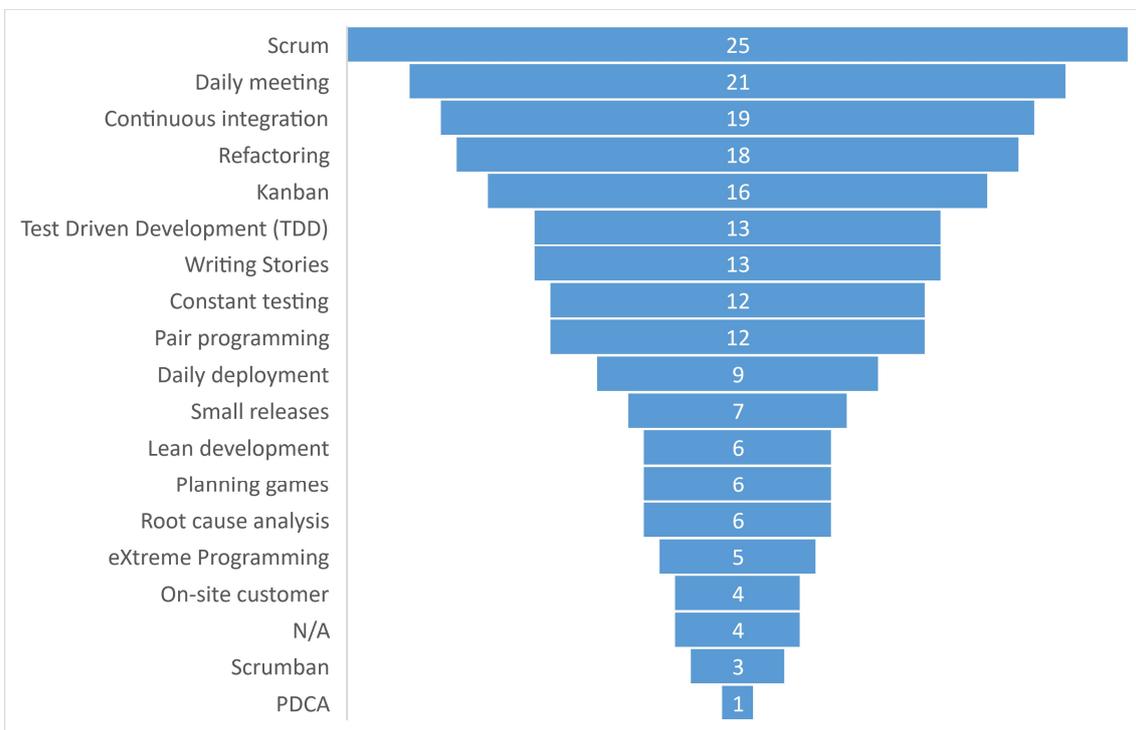

**Figure 8** Level of agility of the organizations

# 4 MODERATOR FACTORS OF SECURITY AND PERFORMANCE VERIFICATION

The S&P moderator factors can be seen as recommendations that must be considered in S&P verification. The next sections present the eight moderator factors (Tables 11-18). First, each of the tables offers an MF and the topics supporting it. Next, it provides information regarding the confirmation of MF and its relevance. The validation and significance of MFs are plotted using a VAS scale in which the dark blue square represents the position of the mean and the area filled with light blue -1 and +1 standard deviation. In the sequence, the tables present the actions used to promote the MF sorted by relevance according to practitioners' opinions. The '#' and '%' represent, respectively, the number and the percentage of participants agreeing to the action contributes to the promotion of MF. Besides, the tables present new actions to promote the MF that were identified in the survey. Finally, it presents the technical literature supporting the MF.

The text presented after each table explains the MF and the topics supporting it. It is important to note that such explanations are based on the case study results (interviews and observations). Therefore, it can represent contextualized results.

## 4.1 MF1: ORGANIZATIONAL AWARENESS OF THE IMPORTANCE OF SECURITY AND PERFORMANCE

Table 11 presents the MF1. It is possible to conclude that the control group agrees with our conclusion stating that *organizational awareness of the importance of S&P* is a factor influencing the S&P verification. Besides, the results plotted on the VAS scale graphically show that this moderator factor has high relevance for S&P verification activities. Thirty-one participants answered about MF1 and the actions to promote this moderator factor. The action '*Keeping programmers well-informed about security and performance*' is the more relevant according to practitioners' opinion (90%) and can be confirmed in technical literature. Moreover, it indicates consensus between practice and academia regarding its relevance.

**Table 11** MF1: Organizational awareness of S&P importance

| MF1: Organizational awareness of security and performance importance | | |
|---|---|---|
| **Need for support from every stakeholder** | High-level managers should support verification activities | ▪ Managers should financially support verification activities<br>▪ Managers should support the purchase of verification tools<br>▪ The project budget should include verification costs |
| | The development team should support the verification activities | ▪ Developers should understand verification as an advantage.<br>▪ It is important to keep a good relationship with developers.<br>▪ Verification activities improve development practices.<br>▪ The development team is responsible for deciding whether to fix faults.<br>▪ Need to know domain and system architecture.<br>▪ Developers may point on what to test.<br>▪ System analysts can assist with the test scenarios prioritization.<br>▪ System analysts can assist in the identification of dependencies between tests scenarios. |
| | Keep customers aware of what is evaluated | ▪ There is no software fault-free software |

| Need for training | Keep the stakeholders aware of the importance of security and performance | ■ Security and performance concerns should not occur only in important deliveries.<br>■ Security and performance concerns should not only occur after the detection of severe failures. |
| | Deal with the absence of well-defined concepts | ■ Standardize the understanding of concepts involving security and performance. |

**MF1 Confirmation (Control group)**

| Moderator factor | Mean | SD | - 1SD | + 1SD | Valid answers |
|---|---|---|---|---|---|
| MF1 | 9,70 | 0,60 | 9,10 | 10,30 | 4 |

**MF1 Relevance (Practitioners' opinion)**

| Moderator factor | Mean | SD | - 1SD | + 1SD | Valid answers |
|---|---|---|---|---|---|
| MF1 | 8,98 | 1,47 | 7,50 | 10,459 | 31 |

| Actions to promote *organizational awareness of security and performance importance* | # | % |
|---|---|---|
| Keeping programmers well-informed about security and performance | 28 | 90% |
| Promoting training | 25 | 81% |
| Informing the customer about the accurate state of software security and performance | 19 | 61% |
| **New actions to promote *organizational awareness of security and performance importance*** | | |
| Simulation of security and performance failures and show business impact | | |
| Regular meetings to discuss security practices | | |
| External audit to mitigate human problems | | |
| Having an ethical hacker would be extremely good for security and creating performance indicators | | |

| Technical literature supporting MF1 | |
|---|---|
| Horký *et al.* (2015) | Report an experiment demonstrating that keeping programmers well-informed about performance can decrease the number of bad decisions influencing system performance. |
| Ferrell and Oostdyk (2010) | Authors emphasize the challenge of security awareness: programmers are not concerned about security because they have a false impression that new development technologies are immune to security problems. |

Security and performance should not be the responsibility of a separate organization department. The overall perception of the importance of the S&P of software systems affects verification activities. Thus, *S&P verification activities require support from every stakeholder*. High-level managers should financially support S&P activities. For instance, they should help purchase verification tools and include the costs of S&P verification in the project budget planning.

*The support of the development team* (e.g., programmers, system analysts, and system architects) is also necessary. They should consider S&P verification as an advantage, understanding that if the verification team reports a failure, this is not against the project. Furthermore, the development team has in-depth knowledge of domain and software architecture. Thus, they can aid decision-making regarding what should be verified, prioritize verification scenarios, and identify the verification scenario dependencies.

*The project customer should also be aware of what is being assessed* and understanding the importance of S&P verification activities. Moreover, customers should know that S&P verification activities do not ensure a fully secure system or a system with no performance issues.

*Continuous training is required* to *keep stakeholders up to date about the importance of S&P of developed software systems*. Unfortunately, organizations generally do not pay proper attention to system security and performance properties, considering S&P verification a waste of resources. Usually, organizations are inclined to be concerned about the S&P of their systems only after having a problem. Another situation where organizations invest more in security and performance assessment is before a major release in which an S&P failure could negatively affect the organization's image.

Besides, *training is also essential to normalize the concepts related to S&P*, furthering the communication between different stakeholders.

### 4.1.1 Actions to promote MF1

It is vital to *keep programmers well-informed about the S&P of the software system they are developing*. Thus, they can understand the consequences of the way they are coding and then improve the coding practices to avoid previous mistakes.

It is essential to *promote training* to the technical staff, the managers, and customers. The programmers should be aware of intrinsic defects of the technologies used in software development and the coding patterns resulting in failures. Thus, they can avoid the use of defective technologies and improve their capability to build failure-free code.

Training is important to managers to understand the S&P verification activities as an investment rather than an expense. In addition, training can alert managers about the problems brought by a software product with poor security and performance (e.g., loss of customers, high infrastructure costs). Thus, managers understand the importance of including S&P verification activities in the development life cycle, reserving budget and time for such activities.

In general, customers do not need to be trained in technical details. However, some of them (e.g., those that participate in software project decisions) can take advantage of training because they can understand the importance of S&P properties. Developing a software product without including S&P verification activities tends to be cheaper and faster. If the customers do not understand the benefits of including S&P verification activities, they will always look for the cheapest option. Therefore, they can agree more easily to the inclusion of S&P verification activities in the project budget.

It is also essential to *keep customers informed about the accurate state of software security and performance* — usually, software development organizations ignore the cost of verification activities in the project budget, offering cheaper software to the customers. Thus, when a failure occurs, the development organization does not notify the customer. Therefore, it thinks the software product is all right and believes that verification activities are a waste of resources. However, the consequences of S&P failures can be catastrophic.

Besides, the survey allows identifying four *new actions to promote the MF1* (Table 11). Further investigations are necessary to confirm the relevance of these recent actions.

## 4.2 MF2: CROSS-FUNCTIONAL TEAMS

Table 12 presents MF2 and the topics supporting this moderator factor. Regarding the MF2 confirmation, it is possible to conclude that the control group agrees that keeping a *cross-functional team* is a factor influencing S&P verification. However, this factor is not unanimous, as the average of the responses is not high, and the standard deviation is wide. It indicates that most control groups agree with MF2, but some participants have low confidence in confirming it.

Twenty-nine participants answered about the relevance of MF2. Looking only at the average opinion, this MF can be considered relevant. However, the significant standard deviation can indicate that practitioners disagree with the average view, understanding that MF2 is more or less relevant than the average opinion.

**Table 12** MF2: Cross-functional team

| MF2: Cross-functional team | |
|---|---|
| **Dependence on specialized verification team** | Need for knowledge on information security policy standards |
| | Need for knowledge on secure development norms |
| | Need for knowledge on digital certification |

| | Identification of software system technologies (Database, Web server, programming language) |
|---|---|
| **Dependence on database team** | Need to request a database restore |
| **Support of infrastructure team** | Configure access permissions |
| | Customize the kernel of the server |
| | Restart the server after a catastrophic failure |
| **Support of legislation experts** | Calculate legal risk |

| MF2 Confirmation (Control group) | | | | | |
|---|---|---|---|---|---|

| Moderator factor | Mean | SD | - 1SD | + 1SD | Valid answers |
|---|---|---|---|---|---|
| **MF2** | 6,82 | 3,45 | 3,36 | 10,28 | 4 |

| MF2 Relevance (Practitioners' opinion) | | | | | |
|---|---|---|---|---|---|

| Moderator factor | Mean | SD | - 1SD | + 1SD | Valid answers |
|---|---|---|---|---|---|
| **MF2** | 7,95 | 2,94 | | 10,89 | 29 |

| Actions to promote the build of a *cross-functional team* | # | % |
|---|---|---|
| Building a team having multiple skills | 23 | 79% |
| Disseminating the view that the verification team is not the enemy but allied | 23 | 79% |
| Stimulating interaction between members of different teams | 18 | 62% |

| New actions to promote the build of a *cross-functional team* |
|---|
| The team should have leaders swapping places (for example, marketing and development). team leaders can get to know limitations, capabilities, and points of view, which can lead to better teamwork and results |
| Highlight the positive results of having a multidisciplinary team |
| Knowing what is problematic in other sectors of the organization |
| Encouraging integration between teams working on similar topics |
| Value verification professionals |
| Select qualified people for the position |
| Invest in the training and qualification of the verification team |
| Apply Scrum |

| Technical literature supporting MF2 | |
|---|---|
| Brucker and Sodan (2014) | Identify the need for teams with different S&P verification activities skills and propose strategies to encourage group information exchange. |
| Williams *et al.* (2010) | Propose a card game called protection poker provides security knowledge sharing, involves the entire development team, and increases software security needs awareness. |
| Johnson *et al.* (2007) | Recommend to consider programmers as allies rather than enemies. Regarding performance, the authors demonstrate how weekly meetings involving performance architects, domain experts, marketing stakeholders, and developers can improve team interactions. |

Verification activities are not performed in isolation by only one team. They require interaction between different teams as well as different skills. In the case studies, a need was identified for experts in security verification, infrastructure, legislation, and databases.

*Security verification experts* are responsible for providing information about security, such as information security policies, security development standards, digital certification, and cryptography. Furthermore, the security verification team should have the required knowledge to perform the fingerprinting step, identifying the technologies for developing the software (database, web server, and programming language).

The infrastructure team's support is outstanding because sometimes the verification teams cannot know how to take some actions. For instance, it may be necessary to allow a specific IP to access the server, make some changes in the operating system's kernel, or restart the server after a catastrophic failure. For example, a participant reported a need to interact with an infrastructure specialist: "*...we have to ask to change the (operating system) kernel because it has a limit in the size (of requests) that can be sent...*"

The verification activities may occasionally affect databases irrevocably. In this case, the *database team* may also aid in the verification activities using its knowledge, for instance, to repair or restore a functional database version.

A need for *legislation expert* support was also identified. Such an expert can be useful in assessing legal risks.

A cross-team interaction is also essential to identify the technologies used for software development and influence the verification results. For example, the verification team observes that the multiple executions of performance test cases decreased response time. Thus, they discovered that the system used the content delivery network (CDN) cache technology to talk to the development team.

### 4.2.1 Actions to promote MF2

*Building a team having many skills* complements the previous practice. The exchange of knowledge between teams is not so useful if every member has the same abilities.

*Disseminating the verification team's view is not the enemy* helps the verification team work together with the other teams. Otherwise, the verification team could have problems in identifying what should be assessed and requesting failure fixes.

Additionally, *stimulating interaction between members of different teams* (i.e., programmers, architects, and requirements analysts) is a way of making explicit each team's capabilities and personal skills. So, when the verification team is faced with a problem that requires specific knowledge to solve, they know whom to ask for help.

Besides, it was also possible to identify eight new actions to promote MF2 that need further investigation.

## 4.3 MF3: SUITABLE REQUIREMENTS

Table 13 presents the MF3 and the topics supporting this moderator factor. The survey result shows the control group's agreement regarding the influence of *suitable requirements* in the S&P verification. Besides, MF3 can be considered relevant according to practitioners' opinions. Moreover, even with a significant standard deviation, the value -1sd is above the scale's midpoint, indicating a consensus about this factor's relevance.

**Table 13** MF3: Suitable requirements moderator factor

| MF3: Suitable requirements | | |
|---|---|---|
| **Lack of well-defined requirements** | Lack of oracle | ▪ Verification activities are used to define system capability. |
| | Generates unnecessary stakeholders interactions | ▪ Generates dependence on functional testing.<br>▪ Generates dependence on the development team. |
| | The verification team does not receive requirements | ▪ The perception of verification team members defines requirements. |
| | Difficulty in determining requirements | |

| The verification team should participate in the requirements phase | The verification team should assess requirements testability |
|---|---|

| MF3 Confirmation (Control group) | | | | | |
|---|---|---|---|---|---|

| Moderator factor | Mean | SD | - 1SD | + 1SD | Valid answers |
|---|---|---|---|---|---|
| MF3 | 8,65 | 2,08 | 6,56 | 10,73 | 4 |

| MF3 Relevance (practitioners' opinion) | | | | | |
|---|---|---|---|---|---|

| Moderator factor | Mean | SD | - 1SD | + 1SD | Valid answers |
|---|---|---|---|---|---|
| MF3 | 8,39 | 2,80 | 5,58 | 11,20 | 31 |

| Actions to promote the building of *suitable requirements* | # | % |
|---|---|---|
| Using techniques to handle security and performance requirements | 25 | 81% |
| Involving the verification team in the requirements phase | 24 | 77% |
| Stimulating the verification team to assess the testability of requirements | 24 | 77% |
| **New actions to promote the building of *suitable requirements*** | | |
| Involving the verification team in all phases of the software life cycle | | |
| The verification team and Product Owner should discuss the specification to identify and adjust any deviations before the specification goes into development. | | |
| The infrastructure team should assess security and performance | | |
| Involving the requirements team in verification activities | | |

| Technical literature supporting MF3 | |
|---|---|
| Harjumaa and Tervonen (2010) Tondel *et al.* (2008) Stephanow and Khajehmoogahi (2017) Weyuker and Vokolos (2000) | These works present challenges regarding S&P verification, making it impossible, ambiguous, or generic: lack of support tools and techniques, unsuitable requirements to target users, lacking requirements, and inaccurate requirement descriptions. |
| McDermott and Fox (1999) Harjumaa and Tervonen (2010) Weyuker and Vokolos (2000) Sindre and Opdahl (2001) Hui and Huang (2012) Jürjens (2002) Bozic and Wotawa (2014) Haley *et al.* (2008) Bulej *et al.* (2017) | A set of proposed techniques and recommendations to handle security and performance requirements can also be identified: misuse cases, SETAM UMLsec, abuse cases, and description of attack patterns indicating a gap between practice and academy, as these techniques are not applied. |

The *lack of security and performance requirements* prevents the verification from fulfilling its original purpose (*i.e.*, assessing whether the software meets its requirements). In the absence of an oracle, it is impossible to know if the verification results are correct. Moreover, *inaccurate requirements overload other teams* (*e.g.*, analysts, architects, and developers) because the verification team must continuously contact them.

In this study's organizations, there were occasionally no written performance requirements that could be used as an oracle. In such cases, the verification activities were not performed to assess whether the software meets its requirements but to evaluate the system's capacity. In other instances, the verification activities were performed based on subjective or imprecise requirements. For example, a participant reported a case where the tests were performed based on a brief description of users' behavior: "*In this system, everyone comes in at 8 in the morning and stays until 10 o'clock. Then they leave the system and come back at lunch*".

The lack of security and performance requirements can be dangerous because the verification team may determine the requirements by their own experience, which may not reflect customer expectations. Furthermore, it was observed that the verification team has some difficulties in deciding S&P requirements.

Participants from two organizations suggest that the *verification team must participate in the requirement phase*, understanding and evaluating their testability.

### 4.3.1 Actions to promote MF3

*Using techniques to identify and represent the S&P requirements* is another action the organizations can perform to improve the way they handle S&P requirements. Using a technique supports the verification team to be more systematic, avoiding misconceptions resulting from subjective decisions. There are different techniques available such as abuse cases, NFR-Framework, Sec-UML. However, it was possible to realize that these techniques did not reach the software development industry.

Regarding the actions, the organizations can *further the involvement of the verification team in the requirements phase*. Thus, the verification team increases their knowledge in the problem domain, favoring the identification of S&P requirements.

Besides, if the requirements phase includes the verification team members, they can provide criticisms regarding how requirements have been represented, improving the specification, and *assessing the testability of the S&P requirements*.

Table 13 also presents the new actions to promote MF3 that could be identified through the survey. Initially, we could determine that '*involving the verification team in the requirements phase*' is crucial to produce formal S&P requirements. However, a participant suggested broader actions stating that it is essential to '*involve the verification team in all phases of the software life cycle.*' Another participant suggested an action with the reverse logic, stating that the *requirements team needs to be involved in the verification activities*. These actions seem to make sense, but they need further researches to improve their understanding and identify their relevance in different contexts.

## 4.4 MF4: SUITABLE SUPPORT TOOLS

Table 14 presents the MF4 and the topics composing it. Besides, it shows that the control group agrees with this MF, allowing us to conclude that our understanding of MF4 is genuine. When asking the practitioners' opinion about the relevance of using suitable tools in S&P verification activities, the results were also satisfactory: the mean is ~7.71. The VAS scale shows that external participants have the perception that MF4 has high relevance to verification activities.

Table 14 also shows thirty-one participants' answers regarding the relevance of the *actions used to promote* MF4 and the suggested *new actions to promote* MF4. Finally, it presents the technical literature supporting MF4.

**Table 14** MF4: Suitable support tools moderator factor

| MF4: Suitable support tools | |
|---|---|
| Support tools decrease the effort of manual activities | |
| **Preference for using free tools** | The high cost of proprietary tools |
| | Avoiding the bureaucracy of buying proprietary tools |
| **Allow the verification team to suggest new tools** | The choice of tools should be discussed among the stakeholders |
| | The tool must be following the team capability |
| **Automated tools generate unsuitable reports** | Too many false positives |
| | Automated tools' report scares the customers |
| | System version |
| | Information about the environment configuration |

| Verification report should include essential information | Performed verification activities | <ul><li>Definition of the performed activities type.</li><li>Identified incidents (with failure evidence, failure explanation, and fix instructions).</li><li>Planned activities but not performed.</li><li>Parts of the system working properly.</li></ul> |
|---|---|---|

| MF4 Confirmation (Control group) | | | | | |
|---|---|---|---|---|---|

| Moderator factor | Mean | SD | - 1SD | + 1SD | Valid answers |
|---|---|---|---|---|---|
| MF4 | 9,22 | 1,41 | 7,80 | 10,64 | 4 |

| MF4 Relevance (Practitioners' opinion) | | | | | |
|---|---|---|---|---|---|

| Moderator factor | Mean | SD | - 1SD | + 1SD | Valid answers |
|---|---|---|---|---|---|
| MF4 | 7,71 | 2,67 | 5,04 | 10,38 | 31 |

| Actions to promote the selection of *suitable support tools* | # | % |
|---|---|---|
| Allowing the technical team to suggest and adopt support tools | 24 | 77% |
| Using tools consistent with the verification team knowledge | 22 | 71% |
| Supporting the use of free tools | 13 | 42% |

| New actions to promote the selection of *suitable support tools* |
|---|
| Providing training to the verification team to enable them to operate the adopted tools (5 participants) |
| Institutionalize the use of tools |
| Using industry best-practice toolsets |
| Support from the tool provider |

| Technical literature supporting MF4 | |
|---|---|
| Thompson (2003)<br>Yee (2006)<br>Johnson *et al.* (2007) | Advocate about the importance of using tools to support verification activities. |
| Guo *et al.* (2010) | Certain types of verification would not be possible without support tools—for instance, long-running tests, significant data volume testing, and concurrent users' tests. |
| Ferme and Pautasso (2017)<br>Shu and Maurer (2007)<br>Ge *et al.* (2006) | Support tools are essential for specific practices or development methodologies, such as continuous integration and agile software development. |
| Johnson *et al.* (2007)<br>Dukes *et al.* (2013) | Automated support tools cannot replace manual verification. Instead, they are complementary practices because current support tools cannot identify some defects. |
| Guo *et al.* (2010)<br>Barbir *et al.* (2007)<br>Kabbani *et al.* (2010)<br>Türpe (2008)<br>Shu and Maurer (2007)<br>Barbir *et al.* (2007)<br>Kim *et al.* (2009)<br>Parveen and Tilley (2008) | These studies highlight the lack of suitable support tools, reporting issues related to the need for integration of different tools, the high cost of proprietary tools, the need for experimental evaluation, the lack of standard-compliant tools, and the lack of support tools targeting specific technologies. |
| Türpe (2008)<br>Luo and Yang (2014)<br>Shu and Maurer (2007)<br>Zhioua *et al.* (2014) | These authors emphasize the excessive false positives generated by current tools and consider this a criterion to choose a suitable support tool. |
| Türpe (2008) | It presents some requirements for useful support tools: in line with the team's capability, idealized site conditions should not be required, and the right problems should be addressed. |

The use of suitable support tools is important in S&P verification activities because it can *decrease the effort of manual activities*. An *inclination to choose free tools* was identified because the acquisition process is faster than the technical team. In adopting proprietary tools, it is necessary to ask the managers for permission, and the price may hinder or impede the buying process.

Moreover, the *verification team should have the autonomy to suggest and adopt new support tools*. The team's capacity should be considered in choosing these tools because the verification team may not have the necessary knowledge to use the tools' advanced features.

Some findings were identified regarding the tools' reports. The first finding regards the excessive number of false positives. In this case, the results can be ignored because it takes substantial effort to analyze each reported failure. Additionally, the tools' report should not be delivered to the customer or the developers. Instead, these results should be investigated and processed by the verification team. Thus, a consistent message can be delivered to the customers or the developers. A participant talks about this: "*...tools generate 'cold' reports. My team and I should analyze and consolidate them, making them more understandable for the users, programmers, and managers.*"

*There is some vital information that a verification report should contain*. Therefore, a suitable tool should generate reports with this information. First, it should provide the system version and configuration information because software or environment settings may require verification re-execution. Second, it is also necessary to provide information on which tests have been performed, define each of them, and report which incidents were detected. Finally, it is needed to provide information to replicate the incident, a possible explanation, and instructions for resolving it in incidents.

Additionally, it is important to inform which tests were planned and not performed (usually owing to deadlines/budget constraints). Finally, it makes the customers more aware of the system's capability and possible production environment failures.

It is also essential to make explicit in the reports the verification activities that did not reveal incidents. It is psychologically positive for the client or developer to know that the system operates correctly in several respects.

### 4.4.1 Actions to promote MF4

It is crucial to *allow the technical team to suggest and adopt support tools*. As intrusion practices constantly evolve, replacing the used tools with a new one or a new version is necessary. The verification team comprises people having the skills to evaluate if a tool is outdated and suggest another.

It is also essential to *use tools consistent with the verification team's knowledge*. An inexperienced verification team using many sophisticated features gets lost and cannot correctly perform the activities. The opposite is also not recommended. A very experienced team using tools with a limited set of features cannot apply all their knowledge and can be considered a waste of resources.

Besides, the *usage of free tools should be encouraged*. This practice allows the technical team to make decisions regarding tools' choices without necessarily involving the management team. It makes selecting and changing the used tools more agile because it avoids the bureaucracy of buying a proprietary tool. However, this action had low relevance, according to the survey's participants. It would mean that there is no preference between using a free or proprietary tool in some organizations. We can hypothesize that a verification tool's cost may be irrelevant to organizations with high economic power or that some organizations have a light purchase tools process, reducing the difficulty of acquiring a proprietary tool.

Besides, the survey results revealed new practices that can promote appropriate tools (Table 14). It is essential to highlight that five of the participants suggested providing training on the tools adopted by an organization. Thus, this action is stronger than the others.

## 4.5 MF5: SUITABLE VERIFICATION ENVIRONMENT

Table 15 presents the MF5 and the topics fostering it. Through the VAS scale, it is possible to identify the control group that confirmed our findings regarding the relevance of having an appropriate environment to perform S&P verification activities. It has a great relevance according to practitioners' opinions.

Thirty-one survey participants answered about the pertinence of the actions used to promote MF5. Thus, Table 15 presents the actions ordered by relevance according to the participants' opinions. Besides, it shows the *technical literature supporting* MF5.

**Table 15** MF5: Suitable verification environment moderator

| MF5: Suitable verification environment | | |
|---|---|---|
| **Verification performed on the unsuitable environment** | Verification and homologation sharing the environment | ▪ Homologation harms verification: need to notify about verification performing; short time window for verification performing.<br>▪ Verification harms homologation: verification kills the server. |
| | Performing verification on a production server | ▪ Real users influence verification results. |
| | Need to execute the tests more than one time | ▪ The network may be overloaded by other people/devices.<br>▪ Cache technology influences the verification results. |
| | Difference between verification environment and production hardware | |
| | Challenges in configuring the application for verification activities | ▪ Creating the required verification data mass.<br>▪ Extra effort to create verification data mass.<br>▪ Data mass creation may depend on other teams. |
| **The verification team should be able to control the environment** | Allowing verification team access/edit verification database | |
| **Virtualization technologies assist in instantiating verification environment** | Making use of virtual machines to simulate the operational environment | |
| | Making use of virtual machines to simulate tests agents | |

| MF5 Confirmation (Control group) | | | | | |
|---|---|---|---|---|---|

| Moderator factor | Mean | SD | - 1SD | + 1SD | Valid answers |
|---|---|---|---|---|---|
| MF5 | 9,87 | 0,18 | 9,68 | 10,06 | 4 |

| MF5 Relevance (Practitioners' opinion) | | | | | |
|---|---|---|---|---|---|

| Moderator factor | Mean | SD | - 1SD | + 1SD | Valid answers |
|---|---|---|---|---|---|
| MF5 | 9,27 | 1,37 | 7,90 | 10,65 | 31 |

| Actions to promote the configuration of a *suitable verification environment* | # | % |
|---|---|---|
| Using virtualization technologies to simulate execution environment | 26 | 84% |
| Keeping the verification team well-informed about used technologies | 22 | 71% |
| Using virtualization technologies to set up tests' agents | 19 | 61% |
| Performing each test case more than once and at a different period to mitigate external influences | 16 | 52% |
| Scheduling the verification activities if it is not possible to instantiate a specific verification environment so that verification should never be performed in parallel with any other activity | 15 | 48% |
| **New actions to promote the configuration of a *suitable verification environment*** | | |
| Using automated verification | | |
| Simulating a defined behavior that constitutes real user behavior | | |
| Using techniques to generate suitable testing data | | |

| Technical literature supporting MF5 | |
|---|---|
| Neto *et al.* (2011) | They use virtualization technologies to support the verification environment due to the financial unfeasibility of using physical machines to compose the verification environment. |
| Arif *et al.* (2018) Kin *et al.* (2015) Gaisbauer *et al.* (2008) | Their concerns relate to the use of virtualization technologies to support verification environments: the estimation of the number of supported virtual machines, the limit of virtual machines, the instability of test trigger response time, and the physical machine overload. |

A suitable environment is essential for verification. In this context, the environment encompasses the configurations of the infrastructure responsible for system operation (e.g., application server and database parameters) and the system's configuration (e.g., the data stored on the database verification activities are performed).

It was observed that the S&P verification occasionally shares the same environment used by other activities. For example, *performance tests were performed on the same server used for user acceptance tests in one organization.* In this case, there was a bidirectional influence. Thus, the performance tests may jeopardize the user acceptance activities because the simulation of many users operating the system causes hardware overloaded. Furthermore, when the system was used for acceptance testing, the performance tests presented unexpected results (*e.g.*, aleatory response time) because it was impossible to know how the users were using the system. In this case, the organization should appropriately schedule the verification activities (performed by the verification team) and the acceptance tests (performed by end-users) so that these two activities never be performed in parallel.

*A production environment is not recommended* because it is difficult to predict its real users' behavior. Thus, the verification results could be misleading if the system is in actual operation.

The local network (or virtual private network) also influences the performance testing results. For instance, if the machine used for performance tests uses the default organization network, the requests and responses may be delayed due to an overload of the network nodes (e.g., routers and sweets) that route them to the server that runs the system.

Some technologies can also influence the results of the verification activities. For instance, using the cache to retrieve data from a database can lead to inaccurate response time test results.

The verification team should resolve network and technology issues by, for example, performing each test case more than once and at different times to mitigate external influences on the test results. For example, one of the participants said: "*It is not possible to rely on the response time of only one scenario execution because there may be interference that impairs the operation of the system. Thus, response time analysis should be performed only after the scenario has been successfully executed three times*".

Another issue regarding the verification environment is the *difference between the hardware configuration used for verification and production time*. In some cases, the hardware used in the production environment is more powerful than the hardware used in verification activities, resulting in false results regarding system performance.

*The difficulty in configuring the system with suitable data for verification activities* was also an issue that was observed. In this case, some participants stated that populating the database with pertinent information is difficult and occasionally requires support from other teams (e.g., database administrators). An alternative to minimizing dependence on different groups is to allow the verification team to access the database's verification activities.

Finally, it was realized that *virtualization technologies are allied with S&P verification*. First, simulate the system execution environment, trying to obtain an environment similar to the production environment. Second, set up the environment through which tests will trigger, for example, creating several virtual machines to simulate simultaneous access to the system.

### 4.5.1 Actions to promote MF5

The participants of the study understand the virtualization technologies as allies of S&P verification. Therefore, it was also possible to confirm the suitability of such technologies in the technical literature. *Furthermore, using a virtual environment at verification time* allows configuring an environment like the production environment. Thus, the results of verification become more realistic.

It is also essential to *keep the verification team well-informed about the technologies used during software building* because such technologies can bias the verification results. For instance, cache technologies can result in different response times according to the software state, masking the system's real performance.

The *virtualization technologies can also simulate testing agents* (*i.e.*, machines from which the tests are trigged). Thus, it is possible to simulate, for example, several users operating the software in parallel, which would be impeditive using real machines.

Different activities use the network infrastructure of the organization (*e.g.*, file transfer, backup routines). These activities can overload the devices (*e.g.*, routers, switches), interfering with the verification results (mostly for performance). Therefore, the *test cases should be performed more than once and at different periods to mitigate external influences*.

Finally, if it is impossible to isolate the verification environment using virtualization technologies, the organization should *create a schedule to perform the verification activities*. This action prevents verification from occurring parallel with other activities (*e.g.*, users performing acceptance testing), avoiding external interferences in the verification results. However, it is a low relevance action as less than half (48%) of the survey's participants understand that it can be used to obtain the appropriate verification environment.

Table 15 also presents the new actions that emerged through the survey and can be used to *configure an appropriate verification environment*. As this is the first time these actions are arising, they still need to be investigated in the future.

## 4.6 MF6: SUITABLE METHODOLOGY

Table 16 presents the MF6 and the topics supporting it. The control group's answers on the VAS scale allow us to conclude that our understanding of a systematic verification methodology's relevance is correct. Besides, the external participants also agree regarding the pertinence of MF6. However, the significant standard deviation indicates that a *systematic verification methodology* does not have high relevance in some contexts.

Besides, Table 16 presents the *actions used to promote* this moderator factor, new actions identified through the survey, and the *technical literature supporting* MF6.

**Table 16** MF6: Suitable methodology moderator factor

| MF6: Systematic verification methodology | | |
|---|---|---|
| **Lack of systematic verification techniques** | Lack of systematic test case selection criteria | ▪ Choice of test cases according to verification team intuition.<br>▪ Test cases based on non-experimental criteria.<br>▪ Difficulty in finding suitable parameters. |
| | Lack of suitable definition of done criteria | ▪ Developers do not know which defects should be fixed. |
| **Organization methodology should be based on previously established standards** | Customize standards to organizational reality | ▪ To align proposed practices with team capacity and development methodology. |

| | The methodology should be adaptable with the used technologies | <ul><li>Discard web threat analysis while verifying embedded systems.</li><li>Discard BD verification while verifying systems without BD.</li><li>Adapt to agile development.</li></ul> |
|---|---|---|
| | Progressive adoption of verification practices | |
| | The methodology must allow evolution | <ul><li>Security and performance needs evolve constantly.</li></ul> |
| **Good methodology requirements** | The need for risk analysis | <ul><li>Identification and prioritization of assets.</li><li>Asset risk level definition.</li><li>To define the security level of each kind of information.</li><li>Managers may provide information on what is critical.</li></ul> |
| | Need for legal authorization | <ul><li>To sign an agreement before starting the verification.</li></ul> |
| | Security and performance verification should be performed after functional verification | <ul><li>To prevent the security and performance verification from identifying functional incidents.</li></ul> |

**MF6 Confirmation (Control group)**

| Moderator factor | Mean | SD | - 1SD | + 1SD | Valid answers |
|---|---|---|---|---|---|
| MF6 | 7,37 | 1,68 | 5,68 | 9,06 | 4 |

**MF6 Relevance (Practitioners' opinion)**

| Moderator factor | Mean | SD | - 1SD | + 1SD | Valid answers |
|---|---|---|---|---|---|
| MF6 | 7,44 | 2,87 | 4,56 | 10,31 | 27 |

| Actions to promote the *systematic verification methodology* | # | % |
|---|---|---|
| Using a proposed methodology and adapting it to the context of the organization | 21 | 78% |
| **New actions to promote the *systematic verification methodology*** | | |
| Modify the company culture at some level by fostering a new methodology | | |
| Search for a methodology aligned with stakeholders needs | | |
| Use appropriately trained testers; avoid using a dopey methodology | | |
| Create processes and revise them according to the proposed methodology and company context | | |

| **Technical literature supporting MF6** | |
|---|---|
| Martin and Xie (2007) | Present the results of an experiment showing the use of a technique increasing the defect detection capability and the coverage of security verification. |
| McDermott and Fox (1999) Alexander (2003) Marback *et al.* (2009) Omotunde and Ibrahim (2015) | Using suitable techniques in different phases of software development (e.g., abuse cases in requirements and modeling, misuse cases, and threat trees in design) promotes identifying defects in the early stages of software development. |
| Omotunde and Ibrahim (2015) Ghindici *et al.* (2006) Brucker and Sodan (2014) | Suggest a combination of techniques to increase the ability to detect different types of defects, e.g., complementing the automated tests with manual reviews, |

| Choliz *et al.* (2015) | Organizations should not develop an entirely new methodology. Instead, it is more suitable to adapt to an existing methodology. |
|---|---|
| Ge *et al.* (2006)<br>Erdogan *et al.* (2010)<br>Sonia and Singhal (2012)<br>Ayalew *et al.* (2013)<br>Wäyrynen *et al.* (2004)<br>Siponen *et al.* (2005) | Inadequacy of existing methodologies in an agile development process and how such methodologies can be adapted to be more agile. |
| Kongsli (2006)<br>Beznosov and Kruchten (2005) | In an agile development process, the lack of documentation and constant refactoring can be impeditive characteristics in implementing current methodologies. |
| Keramati and Mirian-Hosseinabadi (2008) | The demanding activities proposed by real security and performance methodologies can hinder development process agility. In this sense, a proposed metric can be used to measure the agility of a verification methodology. Furthermore, it can be used to evaluate if the adoption of a methodology will impact the agility of the development process. |
| Study (2014) | Some methodologies are challenging to implement. Thus, they should be adapted according to the availability of organizational resources. |
| Study (2014)<br>De Win *et al.* (2009) | Asset identification and risk analysis is an essential requirement of a sound methodology. |

When an organization does not follow a suitable methodology, the verification of S&P is performed in a non-systematic way, impairing its effectiveness and efficiency. For example, if there is no methodology to guide the verification team, the test case selection criteria and the definition of *done* are performed informally, following the tester's intuition.

According to the case study participants, various publications (e.g., pre-defined methodologies, norms, and laws) can be used as the basis for the definition of a methodology in an organization. However, it is not advisable to use these publications verbatim. Instead, it is necessary to understand the recommendations and adapt them to the organization's context, aligning the proposed practices with the practices already used in the organization and with the team's capability. For example, a participant said: "*...knowing that our team is small, I have to work according to our ability, performing the tests for which we have the capacity. I took some courses and could apply other verification activities, but I would need an infrastructure that I do not have.*"

A verification methodology should be adaptable to the technologies used in the development of a software system. For example, it is useless to perform web vulnerability analysis if it is embedded or database verification if it does not hold any data.

Moreover, a verification methodology should allow the increasing adoption of the proposed practices. Thus, the teams can have time to adapt to the new practices. A participant said: "*So, we started using basic open-source tools. Then, we adopted more advanced tools. Thus, using the initial tests, we could understand how we could perform verification and deliver the results to the customers.*"

Moreover, a methodology should evolve because system security and performance should also evolve with time. Regarding security verification, evolution is mandatory, as new invasion techniques are continuously created.

Appropriate points that the methodology should consider were also identified. The first is a risk assessment step, where the assets should be identified, and the criticality level should be assessed. Furthermore, if a third-party company performs the verification activities, the need for legal authorization should be considered.

Finally, the verification methodology should clarify that the security and performance verification activities should be performed after the verification activities targeting the functional requirements; otherwise, the security and performance verification may identify functional failures, contrary to its real purpose.

### 4.6.1 Actions to promote MF6

We could find a single action to promote the use of a systematic methodology (Table 16). This action consists of *choosing a methodology already proposed (e.g., OWASP, Microsoft SDL) and adapting it to the organization's context and particular needs*. The survey result shows that 21 out of 27 participants agree that this action can promote MF6. Besides, the survey allowed the identification of 4 new actions to encourage *a systematic verification methodology*.

## 4.7 MF7: SECURITY AND PERFORMANCE VERIFICATION PLANNING

Another issue regarding S&P verification is related to the planning activity (Table 17). Usually, verification is not well planned, leading to reprioritizing the verification activities and reducing their coverage. Nevertheless, the result summarized in the VAS scale shows that the control group participants agree with the pertinence of MF7. Besides, the 29 valid answers offer that the external participants (practitioners' opinions) understand the planning of S&P verification as an essential moderator factor.

**Table 17** MF7: Security and performance planning moderator factor

| MF7: *[Lack of]* **Security and performance verification planning** | | | |
|---|---|---|---|
| **Security and performance verification requires extra effort** | | | |
| **Insufficient time to perform intended activities** | Need to prioritize verification activities | ▪ Decrease of verification coverage. | |
| | The perception that the verification harms deadlines and costs | ▪ Security development is expensive. | |

| MF7 Confirmation (Control group) | | | | | |
|---|---|---|---|---|---|
|  | | | | | |
| **Moderator factor** | **Mean** | **SD** | **- 1SD** | **+ 1SD** | **Valid answers** |
| **MF7** | 8,1 | 2,20 | 5,89 | 10,30 | 4 |

| MF7 Relevance (Practitioners' opinion) | | | | | |
|---|---|---|---|---|---|
|  | | | | | |
| **Moderator factor** | **Mean** | **SD** | **- 1SD** | **+ 1SD** | **Valid answers** |
| **MF7** | 8,81 | 1,46 | 7,35 | 10,27 | 29 |

| Actions to promote the *planning of security and performance verification* | # | % |
|---|---|---|
| Using a tool to guide the security and performance verification planning | 25 | 86% |
| **New actions to promote the *planning of security and performance verification*** | | |
| Including the security and performance verification activities as part of the development and maintenance cycle | | |
| Having business knowledge helps prioritize the parts of the system that should be evaluated | | |

| Technical literature supporting MF7 | |
|---|---|
| Omotunde *et al.* (2018) | Successful planning can reduce the number of redundant test cases without losing efficiency. |
| Iivonen *et al.* (2010) | Planning ability was recognized as a required skill of good testers. |
| Bozic and Wotawa (2015) | The planning phase can be guided by support tools, decreasing the testers' effort |

The study participants have the perception that *S&P verification activities require additional effort and cost*. Then managers neglect these activities, excluding them from verification planning. A participant presented his opinion about why S&P verification activities are not planned (or were included in the project planning stage): "*... '- How much does it cost to develop a software system?'. '- It costs 300 thousand'. '- And with security?'. '- Well, it depends. So, I should evaluate it. There is a need to have a team performing the verification, and this will have a cost and time impact'. '- So, then leave it for later, for a second version.'...*".

Additionally, while a team of Org1 was performing response time tests, the release time was changed, and some test cases could not be executed. Thus, the team spent more effort reprioritizing the test cases (the planning phase's activity) than performing them.

Moreover, the participants reported that the stakeholders (*e.g.*, managers and customers) have the perception that verification activities can change the delivery time or the cost of a system. However, they do not consider the benefits of these activities. For example, a participant said: "*...every time I talk to someone about testing, about security, or things like that, people always think that it will change the delivery deadline: 'Wow, I need to do it fast.' 'Folks, you are not going to get rework if you do it well.'...*".

### 4.7.1 Actions to promote MF7

It was possible to identify only one practice to promote the MF7. Such action is the *use of tools to guide the planning*, decreasing the effort, and improving the formality of planning activities. 25 out of 29 survey participants agree that this action is relevant to promote security and performance verification.

Additionally, it was possible to identify two new actions to promote MF7. Further investigations are required for understanding the relevance of these new practices and the context they could be applied.

## 4.8 MF8: Encourage Reuse Practices

The reuse of knowledge and artifacts was also identified as a recommendation, bringing more agility to S&P verification activities. Table 18 presents this moderator factor and the actions used to promote it. Besides, it was also possible to confirm the pertinence (control group) and relevance (practitioners' opinion) of MF8.

**Table 18** MF8: Encouraging reuse practices moderator factor

| MF8: Encourage reuse practices | |
|---|---|
| **Reuse of functional test cases** | Functional test cases represent real usage scenarios |
| | It makes security and performance verification more agile |
| **Reuse of previous systems test cases** | Increases the agility of security and performance verification |
| **Use of similar systems to determine requirements** | It helps in defining an oracle |
| **Knowing common defects** | Avoids widely known defects |

| MF8 Confirmation (Control group) | | | | | |
|---|---|---|---|---|---|

| Moderator factor | Mean | SD | - 1SD | + 1SD | Valid answers |
|---|---|---|---|---|---|
| MF8 | 7,8 | 2,43 | 5,36 | 10,23 | 4 |

| MF8 Relevance (Practitioners' opinion) | | | | | |
|---|---|---|---|---|---|

| Moderator factor | Mean | SD | - 1SD | + 1SD | Valid answers |
|---|---|---|---|---|---|
| MF8 | 8,11 | 2,47 | 5,63 | 10,58 | 29 |

| Actions to promote the *reuse of S&P verification practices* | # | % |
|---|---|---|
| Knowing common defects (*e.g.*, vulnerabilities) and using pre-defined test cases to identify the failures caused by these defects | 25 | 86% |
| Reusing the knowledge acquired from other similar systems as a basis for the definition of the requirements | 23 | 79% |
| Reusing functional test cases as they represent real usage scenarios | 19 | 66% |
| Reusing test cases from similar systems adapting parameters | 13 | 45% |
| **New actions to promote the *reuse of S&P verification practices*** | | |
| Creating a base of knowledge of recurring defects | | |
| Mapping vulnerability according to the domain to promote the identification of vulnerabilities applicable to specific situations | | |
| Functional test cases specify what could be added for performance verification | | |
| Design real-time scenarios with production volume data, per hour, per day transaction, Per week. | | |
| Reusing multiple test scenarios is very useful for both professional and runtime scenarios that we can insert in the context of similar new projects | | |

| Technical literature supporting MF8 | | |
|---|---|---|
| Dazhi Zhang *et al.* (2010) Santos *et al.* (2011) | The functional test cases can be reused both as security and performance test cases, bringing benefits such as an increase in the coverage, improvement of failure detection rate, and cost reduction to perform tests and generate suitable testing data set |
| Santos *et al.* (2011) | The reuse of functional testing can bring indirect benefits: an increase in the quality of functional testing because the effort saved in performance testing can be used to improve functional testing; an increase in the diffusion of functional testing owing to their increased importance in the development process |
| Da Silveira *et al.* (2011) | Artifacts generated in the process phase can be reused to automate the generation of performance testing, reducing testing costs |

The *reuse of functional test cases* may be useful in performance tests because they represent real usage scenarios. Moreover, it is possible to adapt the test cases of previous systems' parameters, reducing the construction effort and time.

It was confirmed that *previous similar systems might be used as a basis for defining the requirements*. For instance, a scenario's required response time can be determined based on a similar production system scenario. A participant said: "*The number of concurrent users the system should support is defined by a similar system that is already in production.*"

Finally, it is important to *know common defects* (*e.g.*, vulnerabilities) and to use pre-defined test cases to identify the failures caused by these faults. While talking about penetration tests, a participant mentioned the use of well know cross-site scripting strings (test cases): "*I have a database with more than 350 XSS queries… it is populated with my own knowledge and aggregating other internet databases… For example, OWASP has a database that we can download. Usually, they are the most frequent attacks… I also keep an eye on Exploitdb and vulnerability monitoring platforms. Usually, when they publish an exploit, they also present the XSS query together. So these XSS queries are well known, and it is possible to make them more generic to use in other systems*".

### 4.8.1   Actions to promote MF8

Different services analyze and disclose common vulnerabilities and exposures (CVE). CVEs are defects related to security. In revealing these defects, these services also provide instructions on detecting these defects (test cases). Therefore, the verification team should regularly consult these services *to be aware of common defects and reuse the available test cases*.

Besides, the verification team can *use similar systems as a basis to define S&P requirements*. However, as mentioned before, the S&P requirements may not exist. Thus, using the experience with similar systems already in production can be considered because it allows gaining insight into the new system's needs based on the behavior of actual real users of a similar system.

It is possible to *reuse functional test cases as they represent real usage scenarios*. For instance, it is more appropriate to assess a software feature's response time using real data as input than using randomly generated data.

Finally, it is also possible to reuse the structure of test cases of similar systems, avoiding building them from scratch. However, this action is not relevant according to the opinion of most survey participants (55%), as only 45% state it is appropriate. Thus, it can be considered a low relevance action.

Besides, Table 18 presents the new five new actions we could identify through the survey. As the recent actions identified to promote the other moderator factors, these actions should be further investigated to increase their understanding and relevance in different contexts.

# 5 RELATED WORKS

Considering security and performance as software NFRs, this work aims to understand the state of the practice of NFRs verification. Thus, we delimit the related works to industry studies investigating the verification of NFRs. The difficulty in accessing software development organizations limits the number of such works.

Cruzes *et al.* (2017) report the results of a case study performed in four agile software development organizations. They were concerned about how security testing practices have been done in the agile context. Additionally, they provided some recommendations for practitioners that are in line with our results. First, the teams should stop relying solely on the developers' knowledge regarding security and provide more information about conducting security engineering processes and clear security needs. There is also a danger of external penetration testers do not identify important vulnerabilities due to a lack of domain knowledge. Farther, product owners should have more security awareness, defining and explaining more clearly.

Oyetoyan *et al.* (2018) present the results of action research highlighting developers perceptions while using static application security testing tools: fear of effort to set up a tool, security alerts disrupt regular work, a high number of false positives, the increased cognitive effort to understand tools' messages, limited range of programming languages support, huge technical debts. In addition, they provide implications of the research: one tool is not enough, tools' capability is low, developers have hidden bias regarding static analysis tools report, preference for tools with low integration effort, and developers agree that a tool would improve the security of their product. Therefore, in our work, we advocate the use of appropriate tools.

Werner et al. (2020) present the results of a case study performed on three small software development organizations to identify factors that contribute to a lack of shared understanding of NFRs: the *fast pace of change*, *lack of domain knowledge*, and *inadequate communication*. In our work, we address issues related to the lack of shared understanding of NFRs when we talk about organizational awareness of the importance of NFRs. Additionally, we propose practices that can mitigate this problem. For instance, training can help with inappropriate communication as it establishes a consensus on concepts and facilitates communication.

Behutiye et al. (2020) report the results of a case study on NFRs documentation practices. They conclude that organizations document NFRs using whiteboards and flipcharts, the definition of user stories, and requirements management tools. Their results reinforce our findings, as they also state the importance of NFR documentation for verification activities.

Werner et al. (2021) extends their results presenting practices in the management of NFRs: put a number *metric)* on the NFR, let someone else manage the NFR, write your tool to check the NFR, and put the NFR in source control. They also present challenges: not all NFRs are easy to automate, functional requirements get prioritized over the NFRs, and lack of shared understanding of an NFR. There are several points of intersection with our work. For instance, using metrics can increase the accuracy of the requirements description, helping software verification. Another similar conclusion is that sometimes NFRs cannot be automated assessed.

# 6 THREATS TO VALIDITY

Threats to validity are described following the recommendations of Cruzes and Othmane (2018) and using the quality criteria (Q1–4) and proposed methods (M1–6) to handle them (Lincoln and Guba 2016) (Maxwell 2012).

Credibility (Q1), representing the quality of being convincing or believable, was addressed using rich data/persistent observations (M1) and through data collection using three methods (observation, interviews, and questionnaires), by making notes about what happened, and by verbatim transcripts of what participants said. Furthermore, quotes from the participants were provided.

The transferability (Q2) quality refers to how the results can be generalized to other contexts or settings. This quality is problematic in studies because it is impossible to have many subjects, as was in the present case. However, to improve transferability, an intensive long-term involvement (M2) method was used, whereby the research was conducted on-site, making it possible to have a more accurate contextual perception. Thus, it was possible to provide an in-depth description of the organizations' characteristics and the context in which data were collected.

Regarding dependability (Q3), data stability and reliability over time, and various conditions, the study was conducted in different organizations with participants of multiple profiles. Furthermore, one more case was planned, and it is already scheduled to be performed. Thereby, the results can be triangulated (M3), improving dependability. Additionally, the research protocol is available, making it possible to replicate the study in different contexts.

To avoid researcher bias and improve confirmability (Q4), peer debriefing (M4) was used, exposing the main findings to a research group and discussing their coherence. Furthermore, multiple meetings among the authors were held to discuss the codes. Additionally, a search in the literature was performed to support the conjectures.

The control group survey was used as respondent validation (M5) and member checking (M6) methods. The survey confirmed our understanding of what case study participants said and our conclusions' validity.

Furthermore, the case study was conducted among Brazilian organizations, where Portuguese is the native language. Thus, the participants' quotes reproduced here are translations of what was said. Moreover, the artifacts and codes were initially in Portuguese and translated into English to be presented here. However, this translation does not affect the results reported, as no sentiment/feelings analysis was performed on the answers.

Finally, investigating two NFR's together can be risky. For example, while conducting the case study interviews, it was possible to notice that some practitioners were experts in security OR performance. Thus, even emphasizing that the research context was related to security AND performance, practitioners tend to talk more about their expert property: security OR performance.

# 7 DISCUSSION

Even though the MFs are obvious decisions to ensure that the non-functional verification will run smoothly, we have observed that software organizations did not pay attention to these MFs as needed. Therefore, this paper's first contribution is a detailed list of factors software companies shall pay attention to. Thus, we advocate that these factors should be made explicit and fostered for the software development industry. Furthermore, regarding the application of MFs to other software development activities, further studies shall be conducted.

Aiming to disseminate the MFs and their actions, we made a technical report (http://doi.org/10.5281/zenodo.5129093) to support the spread of this information in the software industry. Such a report provides evidence-based guidance and practical suggestions for practitioners and software organizations to deal with the introduction of S&P activities in their software projects. Besides, all software organizations that took part in the case studies received a feedback presentation. During the presentation, it was possible to notice resistance to implementing all MFs and actions at once for different reasons. Thus, we understand that two strategies could be followed: (1) sort the MFs by relevance (practitioners' opinion) and implement the most relevant with all its actions. After, implement the second more relevant and so on; or (2) implement all MFs, but only the most relevant action. After, implement the second more relevant action and so on. However, the strategy to implement MFs requires further studies.

Next, we present a discussion on each of the MFs, focusing on questions we understand remain open, indicating future research opportunities.

## MF1: Organization awareness of the importance of security and performance

There is a need for a behavioral change of the people involved in the development process (including users and customers) to address this MF. Every stakeholder should be aware of the extra effort and cost needed to build software with suitable security and performance. They also should be aware of the consequences of insecure and unsuitable performance software. In this way, they can decide about the appropriate level of security and performance of their software.

We understand that this MF's difficulty is to bring information to each stakeholder at the appropriate level. So, for instance, while technical staff better understand *SQL injection* and *buffer overflow*, business people better understand things like *damaging the organization's image* and *losing customers.* That is, talking about technical stuff to technical staff and talking about business to business people.

Although this work presents actions to improve S&P awareness, it is necessary to understand better how to apply these actions in software development organizations. Thus, the following questions can be an inspiration for further research: *How to improve awareness of the importance of S&P according to stakeholders' profiles?*

## MF2: Cross-Functional teams

Software development is an intellectual activity that depends on different skills. This MF is related to this statement and is also in line with what is proposed by agile methodologies. For instance, the concept of squads in which an organization is divided into small multidisciplinary groups with specific objectives. Considering this concept, we can say that much knowledge is greater than a specialist's knowledge. However, the major difficulty is that there is still much to learn on cross-functional teams, especially in non-traditional agile roles, such as security expert, performance expert, and others: What are the skills needed for S&P verification? *How to build multidisciplinary teams balancing different skills?*

## MF3: Suitable requirements

The lack of requirements or the inappropriateness of them is a recurrent problem in software development. S&P verification is not an exception. In this sense, we concluded that this MF is extremely important to be considered when implementing S&P verification activities. Besides, the identified actions related to MF3 are valuable. The challenges associated with MF3 are the lack of knowledge on how to identify and represent S&P requirements. Thus, the questions presented are not a novelty: *How to identify and describe security and performance requirements?*

## MF4: Suitable support tools

Tools are essential for S&P verification as some testing techniques are not viable without using them. For instance, it is unfeasible to simulate hundreds of users accessing a service (performance) or a brute force attack using a 100-thousand-word dictionary (security) without a tool's support. So, knowing the need for a support tool, the question is: *What criteria should be used to select a suitable support tool for S&P verification?*

## MF5: Suitable verification environment

MF5 can be divided into two parts: (1) the configuration of the infrastructure used for system verification and (2) the configuration of the system under testing (SUT). Regarding the configuration of the infrastructure, we concluded that isolation is the core question. However, virtualization technologies meet the need to create an isolated environment. Nonetheless, it is still necessary to disseminate the idea that verification activities should be performed remotely, without external interferences. Regarding the SUT configuration, we realized that the challenges are configuring the SUT initial state that allows the verification to start (*e.g.*, suitable testing data) and reset this state quickly after a testing battery. Therefore, the main open question is: *What techniques can be used to configure a SUT before starting S&P verification?*

## MF6: Systematic verification methodology

While we identified organizations to participate in the case study, we often received answers such as "we do not perform S&P verification because we do not know what to do." However, a methodology works as a guide defining what should be done. Besides, a methodology supports the S&P verification planning since knowing what can be done. Therefore, it is possible to determine what will be done. Section 4.6 recommended using an existing methodology but adapted to the organization's reality and the software being developed. Thus, the question that we concluded is still open is: *How to select a suitable S&P methodology for an organization/project?*

## MF7: Security and performance verification planning

Planning is an important activity in every engineering area, including software engineering and S&P verification. However, we could observe that there is a perception that S&P activities hurt the cost and delay the deadline of a software project. Therefore, S&P verification planning is neglected. However, we could not identify the harm caused by not planning such verification activities in our research. Thus, MF1 could help with this issue. Besides, we understand that answering the following question can help the S&P verification planning area: *What are the planning strategies for S&P verification?* and *how to choose a suitable strategy?*

## MF8: Reuse practices

Reuse is a consolidated area of software engineering, with the benefit of avoiding rework and improving development agility. As mentioned in Section 4.8, S&P verification can benefit from reuse practices, not only from the reuse of artifacts but also from the knowledge acquired when developing previous similar systems. Thus, the question is to know what can be reused and how to reuse: *what can be reused to improve S&P verification, and how can it be reused?*

## 8 Conclusion

We presented eight S&P moderator factors that emerged from the results of a multiple-case study with four software development organizations. Besides, the results were strengthened through rapid reviews and a survey. These moderator factors represent topics a software development organization should consider implementing or improve S&P verification activities. Additionally, we also present actions to promote the moderator factors, helping software development organizations to reach them. These findings showed that software development organizations should (1) promote the awareness of the importance of the S&P, (2) keep a cross-functional verification team, (3) produce precise S&P requirements, (4) make use of suitable S&P verification tools, (5) configure an adequate S&P verification environment, (6) use an S&P verification methodology, (7) plan the S&P verification activities, and (8) encourage reuse practices.

The moderator factors and their actions were evaluated using a systematic research methodology using the technical literature and practitioner's opinion. Therefore, we are confident about using these findings as practical guidance to introduce or improve the S&P verification in software development organizations.

## Acknowledgment

We much appreciate the support of participating organizations and CAPES. Prof. Travassos is a CNPq Researcher (grant number 304234/2018-4). This work was partially supported by the SoSAgile project: Science of Security in Agile Software Development, funded by the Research Council of Norway (grant number 247678).